\documentclass[conference]{IEEEtran}
\IEEEoverridecommandlockouts
\usepackage{cite}
\usepackage{amsmath,amssymb,amsfonts}
\usepackage{algorithmic}
\usepackage{graphicx}
\usepackage{textcomp}
\usepackage{xcolor}
\def\BibTeX{{\rm B\kern-.05em{\sc i\kern-.025em b}\kern-.08em
    T\kern-.1667em\lower.7ex\hbox{E}\kern-.125emX}}

\usepackage{amsmath, amscd, amsfonts}
\usepackage{mathtools}
\usepackage{graphicx}
\usepackage{textcomp}
\usepackage{xcolor}
\usepackage{color}
\usepackage{listings}
\usepackage{multirow}
\usepackage{makecell}
\usepackage{subfigure}
\usepackage{xspace}
\usepackage{enumitem}
\usepackage{balance}
\usepackage{algorithmic}
\usepackage[ruled,linesnumbered,vlined]{algorithm2e}
\usepackage{amsthm} 
\usepackage{caption}

\usepackage{hyperref}
\usepackage{etoolbox}
\usepackage{booktabs}
\usepackage{microtype}
\usepackage{textcomp}
\usepackage{pgfplots}
\usepackage{tikz}
\usepackage{xcolor}
\usetikzlibrary{arrows.meta}
\pgfplotsset{compat=1.18} 
\definecolor{safe-black}{RGB}{0,0,0}
\definecolor{safe-olive}{RGB}{0,73,73}
\definecolor{safe-teal}{RGB}{0,146,146}
\definecolor{safe-pink}{RGB}{255,109,182}
\definecolor{safe-peach}{RGB}{255,160,110}
\definecolor{safe-plum}{RGB}{73,0,146}
\definecolor{safe-cerulean}{RGB}{0,109,219}
\definecolor{safe-lavender}{RGB}{182,109,255}
\definecolor{safe-sky}{RGB}{109,182,255}
\definecolor{safe-baby}{RGB}{182,219,255}
\definecolor{safe-brick}{RGB}{146,0,0}
\definecolor{safe-brown}{RGB}{146,73,0}
\definecolor{safe-orange}{RGB}{219,209,0}
\definecolor{safe-green}{RGB}{36,255,36}
\definecolor{safe-yellow}{RGB}{255,255,109}

\usepackage{listings}
\usepackage{comment}
\usepackage{bm}

\definecolor{magenta4}{rgb}{0.5625,0,0.5625}
\definecolor{green4}{rgb}{0,0.5625,0}
\definecolor{orange4}{rgb}{0.98,0.31,0.09}

\def\BibTeX{{\rm B\kern-.05em{\sc i\kern-.025em b}\kern-.08em
    T\kern-.1667em\lower.7ex\hbox{E}\kern-.125emX}}

\newcommand{\algnameshort}{pdGRASS\xspace}
\newcommand{\pdgrass}{pdGRASS\xspace}
\newcommand{\numgraphs}{18\xspace}
\newcommand{\avgspeedupalphatwo}{$8.8\times$\xspace}

\newcommand{\avgspeedupalphaten}{$3.9\times$\xspace}
\newcommand{\iterfe}{\ensuremath{iter_{fe}}\xspace}
\newcommand{\iterpd}{\ensuremath{iter_{pd}}\xspace}

\newcommand{\avgiteralphatwo}{$1.2\times$\xspace}

\newcommand{\avgiteralphaten}{$1.8\times$\xspace}

\newcommand{\avgparallelspeeduptwo}{$11.3\times$\xspace}
\newcommand{\avgparallelspeedupfive}{$13.7\times$\xspace}
\newcommand{\avgparallelspeedupten}{$14.8\times$\xspace}

\newcommand{\pgrass}{pGRASS\xspace}

\newcommand{\fegrass}{feGRASS\xspace}
   
   \newcommand{\prblshort}{GSS\xspace}
   \newcommand{\numthreads}{32\xspace}
      \newcommand{\cutoff}{1E5\xspace}
         
   \newcommand{\innerpar}{inner\xspace}
   \newcommand{\outerpar}{outer\xspace}

   \newcommand{\lca}{\ensuremath{\operatorname{LCA}}\xspace}
   \newcommand{\dist}{\ensuremath{\operatorname{dist}}\xspace}
   \newcommand{\su}{\ensuremath{S_{u, \beta}}\xspace}
   \newcommand{\sv}{\ensuremath{S_{v, \beta}}\xspace}
   \newcommand{\defn}[1]{{\textit{\textbf{\boldmath #1}}}\xspace}
   \SetKwFor{ParallelFor}{parallel for}{do}{endfor}
\newtheorem{theorem}{Theorem}
\newtheorem{lemma}[theorem]{Lemma}

\newtheorem{definition}[theorem]{Definition}

\newcommand{\figref}[1]         {Figure~\ref{fig:#1}}

\newcommand{\defref}[1]        {Definition~\ref{def:#1}}
\newcommand{\tabref}[1]        {Table~\ref{tab:#1}}
\newcommand{\secref}[1]         {Section~\ref{sec:#1}}
\newcommand{\lemref}[1]           {Lemma~\ref{lem:#1}}
\newcommand{\lemreftwo}[2]       {Lemmas~\ref{lem:#1} and~\ref{lem:#2}}

\newcommand{\figreftwo}[2]      {Figures \ref{fig:#1} and~\ref{fig:#2}}

\newcommand{\para}[1]{\smallskip\noindent\textbf{#1.}}
\begin{document}

\title{pdGRASS: A Fast Parallel Density-Aware Algorithm for Graph Spectral Sparsification
}

\author{\IEEEauthorblockN{Tiancheng Zhao}
\IEEEauthorblockA{
\textit{Georgia Institute of Technology}\\
tzhao350@gatech.edu}
\and
\IEEEauthorblockN{Zekun Yin}
\IEEEauthorblockA{
\textit{Shandong University}\\
zekun.yin@sdu.edu.cn}
\and
\IEEEauthorblockN{Huihai An}
\IEEEauthorblockA{
\textit{Shandong University}\\
202335297@mail.sdu.edu.cn}
\and
\IEEEauthorblockN{Xiaoyu Yang}
\IEEEauthorblockA{\textit{China University of Petroleum-Beijing} \\
yangxiaoyu@student.cup.edu.cn}
\and
\IEEEauthorblockN{Zhou Jin}
\IEEEauthorblockA{
\textit{Zhejiang University}\\
z.jin@zju.edu.cn}
\and
\IEEEauthorblockN{Jiasi Shen}
\IEEEauthorblockA{\textit{Hong Kong University of Science and Technology}\\
sjs@cse.ust.hk}
\and
\IEEEauthorblockN{Helen Xu}
\IEEEauthorblockA{
\textit{Georgia Institute of Technology}\\
hxu615@gatech.edu}
}

\maketitle

\begin{abstract}

    Graph Spectral Sparsification (\prblshort) identifies an ultra-sparse subgraph, or \emph{sparsifier}, whose Laplacian matrix closely approximates the spectral properties of the original graph, 
    enabling substantial reductions in computational complexity for computationally intensive problems in scientific computing.
    The state-of-the-art method for efficient GSS is \emph{\fegrass},
    consisting of two steps: 1) spanning tree generation and 
    2) off-tree edge recovery. 
    However, \fegrass suffers from two main issues: 
    1) difficulties in parallelizing the recovery step for strict data dependencies, 
    and 2) performance degradation on skewed inputs, often requiring multiple passes to recover sufficient edges.

    To address these challenges, we propose \defn{parallel density-aware Graph Spectral Sparsification} (\algnameshort), a parallel algorithm that organizes edges into disjoint \emph{subtasks} without data dependencies between them, enabling efficient parallelization and sufficient edge recovery in a single pass. We empirically evaluate \fegrass and \algnameshort based on 1) off-tree edge-recovery \emph{runtime} and 2) sparsifier quality, measured by the \emph{iteration count} required for convergence in a preconditioned conjugate gradient (PCG) application.
    The evaluation demonstrates that, depending on the number of edges recovered, \algnameshort achieves average speedups ranging from \avgspeedupalphaten to \avgspeedupalphatwo. The resulting sparsifiers also show between \avgiteralphatwo higher and \avgiteralphaten lower PCG iteration counts, with further improvements as more edges are recovered. Additionally, \algnameshort mitigates the worst-case runtimes of \fegrass with over $1000\times$ speedup. These results highlight \algnameshort's significant improvements in scalability and performance for the graph spectral sparsification problem.
    
\end{abstract}


\section{Introduction}\label{sec:intro}

Graph spectral sparsification (\prblshort) aims to construct a
\defn{sparsifier}, which is a nearly-linear-sized subgraph that can approximate the spectral information (i.e., eigenvalues and eigenvectors) of the original graph.
Researchers have investigated this problem in both theory~\cite{cohen2014solving, koutis2010approaching, spielman2011graph, spielman2014nearly} and practice\cite{Feng:2016, feng2020grass, liu2021fegrass, zhang2020sf, zhao2024CSP, Bai2024Date}. 
Using the sparsifier rather than the original graph facilitates near-linear-time algorithms for a
wide variety of computationally-intensive matrix and graph applications such as
scientific computing~\cite{spielman2014nearly}, max-flow problems in
graphs~\cite{christiano2011electrical,
  koutis2014approaching,spielman2011spectral, spielman2014nearly}, data
mining~\cite{peng2015partitioning}, social network analysis~\cite{TS16}, solving
large systems of equations~\cite{ZhaoWaFe17}, machine
learning~\cite{defferrard2016convolutional}, and computer-aided design (CAD) for
very large-scale integration (VLSI)~\cite{Feng:2016,zhao2017spectral}.  Due to
the heavy computational complexity of these applications, graph sparsification
is necessary to solve large-scale problems in a reasonable timeframe.
Unfortunately, most existing implementations of algorithms for \prblshort are
sequential and stop short of parallelization.

\para{Edge recovery in sparsification} The state-of-the-art method for
\prblshort is \fegrass~\cite{liu2021fegrass} and \pgrass ~\cite{pgrass} (a parallel implementation of \fegrass but not open-sourced), which performs two key steps: 1)
spanning-tree generation and 2) recovering ``dissimilar'' spectrally-critical
off-tree edges that were not chosen in the first phase but are necessary for
improving the approximation quality. The main challenge in designing parallel algorithms
for \prblshort is the \emph{off-tree edge recovery} phase, as spanning-tree
generation has well-known optimized solutions.

\para{Challenges to parallelization from data dependencies}
An off-tree edge can be recovered only if it is not similar to all previous recovered edges.
In other words, each iteration is inherently dependent on the results of the previous iterations, imposing a sequential constraint on the overall process.
The state-of-the-art \emph{parallel} implementation for \prblshort is
\pgrass~\cite{liu2021pgrass, liu2023pgrass}, which parallelizes over edges in batches but incurs \emph{excess work}.
Edges in the same batch can be similar edges. 
Specifically, if there are $p$ threads, \pgrass performs similarity check on the next $p$ edges simultaneously while there can be redundant computation, which is unavoidable for the correctness of the parallel algorithm.

\emph{Since there is no available implementation for \pgrass, we compare our
  algorithm 
  directly with \fegrass.}

\para{Worst-case inputs}
The off-tree edge recovery phase only stops when the algorithm recovers a fixed number of edges of $\alpha|V|$, where $\alpha$ is a predefined ratio that specifies the proportion of edges to be recovered (default value is 2\%)
and $|V|$ is the number of vertices.
For a small $\alpha$ such as 0.01, one pass through the off-tree edges
is sufficient to recover enough edges, but when $\alpha$ is set to
a moderately larger value, such as 0.05, the \fegrass strategy may
require many passes through the off-tree edges to recover enough edges
to meet the required threshold.
In an extreme case, on the \textit{com-Youtube} graph in part \ref{sec:eval},
we found that \fegrass requires over $6000$ passes to recover
$0.02|V|$ edges.
It reveals that algorithm based on vertex cover degrades severely on certain special graph structures.

\para{Density-aware graph spectral sparsification}
To address these issues, we propose \defn{parallel density-aware
Graph Spectral Sparsification}
(\algnameshort), a fast and accurate parallel algorithm for \prblshort.
Like \fegrass and \pgrass, \algnameshort follows a two-step framework but introduces two key modifications:
1) changing the edge-similarity condition to a \emph{``strict''} condition
to recover more edges in one pass for various graph structures,
which gives rise to 2) a careful parallelization strategy based on division of
the entire problem into \emph{independent ``subtasks''}.
The proposed subtask-level parallelization significantly reduces
the data dependencies between edges
in \fegrass and \pgrass through subtask partitioning.
Furthermore, we find that the strict condition enables
\algnameshort to complete recovery in a single pass
by retaining more edges.

\para{Mixed parallel strategy}
We empirically evaluate several parallelization
strategies for \algnameshort and propose a \defn{mixed} approach that
exploits parallelism both \emph{within} and \emph{across} subtasks.
This is necessary because real-world graphs, characterized by a few
high-degree and many low-degree vertices, lead to skewed subtask sizes.
Parallelizing only along one dimension would leave performance on the table and unnecessarily serialize the computation.
Meanwhile, a mixed parallel strategy can adapt the ratio of the two parallel strategies based on the number of available cores, achieving better speedup.

\begin{figure}[t]
  \centering
  \includegraphics[width=0.62\linewidth]{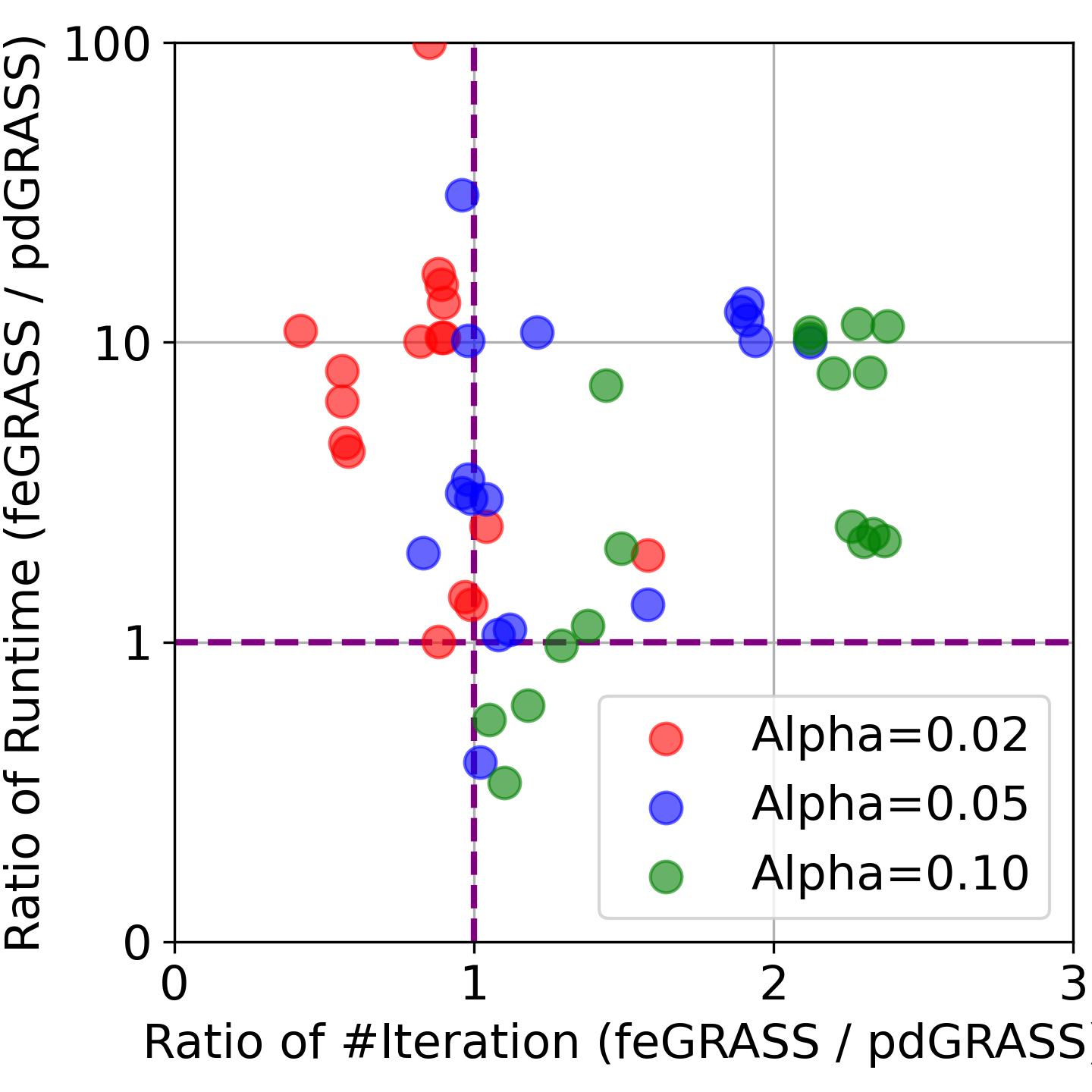}
  \caption{Relative off-tree-edge recovery time and PCG iteration count
    (\fegrass / \algnameshort) for 18 test graphs.
    A number above 1 in either metric means that \algnameshort improves along that metric compared to \fegrass.}
\label{fig:trade-off}
\end{figure}

\para{Results summary}
We evaluate the state-of-the-art serial algorithm \fegrass\footnote{We compare with \fegrass because no open-source implementation of \pgrass is available.}\cite{liu2021fegrass} and the proposed \algnameshort based on:
1) their off-tree edge recovery \emph{runtime}, and 2) the \emph{iteration count} required for convergence in a downstream preconditioned conjugate gradient (PCG)\cite{barrett1994templates} solver.
~\figref{trade-off} summarizes the results in terms of runtime and
PCG iteration count on a suite of \numgraphs input graphs.
As $\alpha$ increases and more edges are required to be recovered, the data points collectively shift toward the upper right direction,
reflecting consistent improvements of \pdgrass over \fegrass in both run-time efficiency and sparsifier quality as shown in \figref{trade-off}.
Finally, as detailed in~\secref{eval}, \algnameshort achieves strong parallel scaling on both uniform and skewed inputs, enabled by its mixed parallel strategy across and within subtasks. On \numthreads threads, \algnameshort attains an average parallel speedup of \avgparallelspeeduptwo for $\alpha = 0.02$, and further improves to \avgparallelspeedupfive and \avgparallelspeedupten for $0.05$ and $0.10$.


\section{Preliminaries}\label{sec:prelim}

This section presents background and current algorithms for graph spectral
sparsification problem.  
We introduce key concepts such as the graph Laplacian matrix and then review with the
efficient serial algorithm \fegrass~\cite{liu2021fegrass} and \pgrass~\cite{pgrass}.

\subsection{Graph Spectral Sparsification}

The Graph Spectral Sparsification (\prblshort) problem takes
a weighted undirected graph $G = (V, E, w)$ as input and generates an ultra-sparse subgraph (also called a \defn{sparsifier}) $P$ as output which is ``spectrally similar'' to the original graph $G$.

We first define the \defn{Laplacian matrix} of a graph $G$, denoted by
$L_G \in \mathbb{R}^{|V| \times |V|}$.
\begin{equation}
L_G(i, j)=
\begin{cases}
-w_{i, j}, & (i, j) \in E\\
\sum\limits_{(i, k) \in E} w_{i, k}, & i = j\\  \label{eq:LG}
0, & \text{otherwise.}
\end{cases}
\end{equation}
A graph $P$ is \defn{$\sigma$-spectrally similar} to a graph $G$ if for any
$u \in \mathbb{R}^{|V|}$,
\begin{equation}
    \frac{1}{\sigma}u^TL_Pu \leq u^TL_Gu \leq \sigma u^TL_Pu
\end{equation}
where $L_G$ and $L_P$ are the Laplacian matrices of graphs $G$ and $P$,
respectively~\cite{batson2013spectral}. If graphs $G$ and $P$ are
$\sigma$-spectrally similar, then $\kappa(L_G, L_P)\leq \sigma^2$, where
$\kappa(L_G, L_P)$ denotes the relative condition number\footnote{The condition
  number of two matrices $A$ and $B$ is defined as $\kappa(A, B) =
  \lambda_{\text{max}}(A, B)/\lambda_{\text{min}}(A, B)$ \label{eq:kappa} where
  $\lambda_{\text{max}}$ and $\lambda_{\text{min}}$ denote the largest and
  smallest nonzero generalized eigenvalues, respectively.}  of their
Laplacian matrices.

\subsection{\fegrass Algorithm~\cite{liu2021fegrass}}

\fegrass is an efficient serial algorithm for \prblshort, which consists of
two main steps: 1) spanning-tree generation based on the \textit{Effective Weight}
of edges and 2) off-tree edge recovery based on the \textit{Resistance Distance}
of paths in the spanning tree.

\begin{definition}[Effective weight]
  The \defn{effective weight}~\cite{liu2021fegrass} $W_{\mathrm{eff}}$ of an
  edge $e=(i, j)$ is defined as follows:
  \begin{equation}
      W_{\mathrm{eff}}(e) = w_{u, v} \times \frac{\log (\max \{\operatorname{deg}(u), \operatorname{deg}(v)\})}{\operatorname{dist}_G(root, u)+\operatorname{dist}_G(root, v)}
      \label{eq:weff}
  \end{equation}
    where $G$ denotes the original graph and $root$ is the vertex with the
    maximum degree in $G$ and $\text{dist}_G(\cdot)$ refers unweighted distance,
    which can be computed by Breadth-First Search(BFS).
\end{definition}

\begin{definition}[Resistance distance]
  Given a spanning tree $T$, the \defn{resistance distance}
  $R_{T}$ (paraphrased\footnote{In the original
    \pgrass~\cite{liu2021pgrass, liu2023pgrass} paper, the algorithm partitions
    the spanning tree so it can run Tarjan's offline least common ancestor (LCA)
    algorithm~\cite{GabowTa83}. In this paper, we dynamically compute the LCA,
    so we do not need to partition the spanning tree.}  from~\cite{liu2021fegrass})
  of an edge $e = (u, v)$ is defined as
  \begin{equation}
      R_{T}(u, v) = \operatorname{dist}_{re}(u, \operatorname{LCA}_T(u, v)) +
  \operatorname{dist}_{re}(v, \operatorname{LCA}_T(u, v)),
    \label{eq:RT}
  \end{equation}
  where $\operatorname{dist}_{re}$ is distance of $u, v$ based on resistant weight
  $W_{re}$ and $W_{re}(e) = 1/w(e)$.
\end{definition}

After spanning-tree generation, each edge in the original graph belongs to
either the \defn{tree edge} set and \defn{off-tree edge} set.
The next step is off-tree edge recovery based on the concept of ``resistance distance'' from the theory of electrical resistance networks~\cite{babic2002resistance,spielman2011graph} which determines spectrally criticality.

Intuitively, two off-tree edges are ``similar''
if they have ``similar positions on the spanning tree''.
Formally, similarity is determined by vertex neighborhoods of the endpoints of an off-tree edge, which can be computed with BFS process.


\begin{figure}[t]
  \centering
  \includegraphics[width=.6\linewidth]{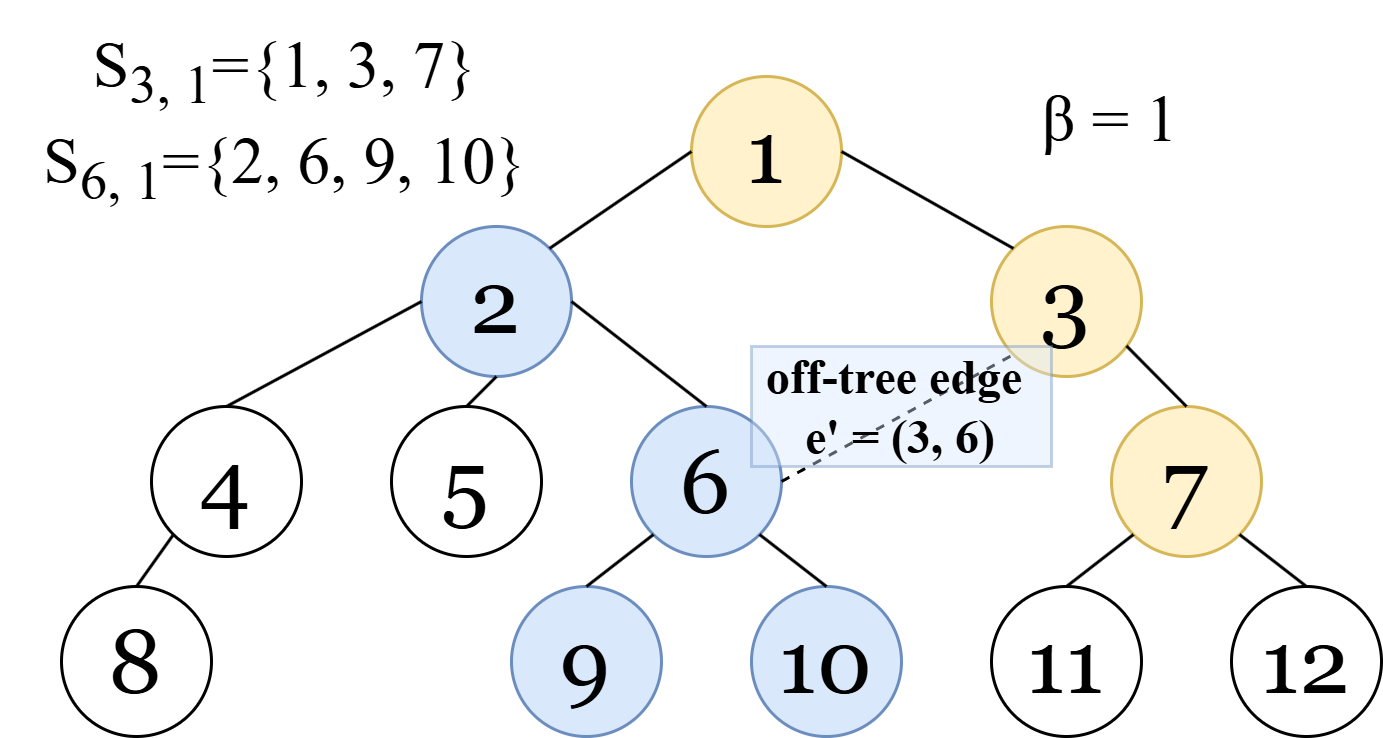}
  \caption{An illustration of the similarity check for a recovered off-tree edge
    $e = (3, 6)$ with BFS step size $\beta = 1$.
    The blue vertices represent $S_6$ and yellow vertices represent $S_3$.
    }
  \label{fig:similar-edge}
\end{figure}

\begin{definition}[BFS step size and vertex neighborhoods]\label{def:bfs-step}
  Let $e = (u, v)$ be a recovered off-tree edge and let $T$ be the spanning tree
   from the first step. 
  The corresponding \defn{BFS step size} $\beta$ is defined as a small constant $c$
  (default value is 8) in the feGRASS algorithm and let $S_{u, \beta}$
  (resp., $S_{v, \beta}$) be the \defn{$\beta$-hop
  neighborhood} of vertex $u$ (resp., vertex $v$).
  That is, $S_{u, \beta}$ is a set containing all vertices with distance
  at most $\beta$ from $u$ (i.e., all vertices visited in $\beta$ BFS iterations from $u$):
\begin{equation}
   S_{u, \beta} = \{ v: \text{dist}(u, v) \le \beta, \beta = c\} \label{eq:Su}
\end{equation}
\end{definition}

\begin{definition}[Similiar edge (Loose Definition)]
\label{def:similar-edge-loose}
  Suppose we have a spanning tree $T$, an off-tree edge $e = (u, v)$, and
  BFS step size $\beta$ and get two $\beta$-hop vertex neighborhood sets
  $S_{u, \beta}, S_{v, \beta}$.

  An off-tree edge $e' = (u', v')$ is \defn{similar} to off-tree edge $e$ if
    \begin{equation}
        (u' \in S_{u, \beta} \vee v' \in S_{v, \beta})
        \vee
        (u' \in S_{v, \beta} \vee v' \in S_{u, \beta}).
        \label{eq:cond-loose}
    \end{equation}
\end{definition}

The process of recovering similar edges can be interpreted as a \textbf{Vertex
  Cover} problem and we can derive an equivalent
formulation of the condition \ref{eq:cond-loose} stated above as follows.
    \begin{equation}
        (u' \in S_{u, \beta} \cup S_{v, \beta})
        \vee
        (v' \in S_{u, \beta} \cup \in S_{v, \beta}).
        \label{eq:cond-loose-2}
    \end{equation}

The size of the resulting subgraph in the \fegrass algorithm is constrained by the input parameter $\alpha$, determining the number of recovered edges as $\alpha|V|$.
The generated subgraph contains $|V| - 1 + \alpha |V|$ edges in total.
If \fegrass fails to recover $\alpha|V|$ edges after one pass, the algorithm will repeat the off-tree-edge recovery process on the remaining off-tree edges until it recovers $\alpha|V|$ edges.

\subsection{Parallelizing the \fegrass Algorithm}

Subsequent work introduced \pgrass~\cite{liu2021pgrass}, a parallel algorithm
based on \fegrass, which focuses on parallelizing the off-tree edge recovery
step because the other steps have standard parallel solutions.  First, the
algorithm computes the effective resistance of every edge in the original graph
using the spanning tree, which can be parallelized using divide-and-conquer to
divide the spanning tree into subtrees~\cite{liu2021pgrass} and applying
Tarjan's offline LCA algorithm~\cite{GabowTa83}.

At a high level, \pgrass parallelizes the seemingly-sequential edge-similarity
checks by dividing the off-tree edges upfront into sequential blocks such that
parallel threads process edges within each block in parallel.  Note that the
algorithm may perform \emph{excess work} because it may process edges that are
similar to earlier edges in the block in parallel.  After all edges in a
block have been processed, \pgrass performs a serial pass through all edges in
the block to check whether they should truly be recovered, or if they should
have been skipped due to parallel edges within the same block.

\subsection{Analysis method}

In this paper, we analyze parallel algorithms in the \defn{work-span model}~\cite[Chapter~27]{CLRS}.  The \defn{work} is the total time to execute
the entire algorithm in serial. The \defn{span} is the longest serial chain of
dependencies in the computation.  In the work-span model with binary forking, a
parallel for loop with $k$ iterations with $O(w_i)$ work and $O(S_i)$ span in
the $i$-th iteration has $O(\sum\limits^k_{i=0} w_i)$ work and
$O(\log(k) + \max\limits_{i=0}^k S_i)$ span.


\section{Parallel Density-Aware Graph Spectral Sparsification}\label{sec:algorithm}

This section proposes \defn{parallel density-aware Graph Spectral
Sparsification} (\algnameshort), an efficient algorithm for \prblshort that
resolves the drawbacks of \fegrass:
1) loose approximation and 2) redundant work during parallelization.

The \algnameshort algorithm modifies the off-tree edge recovery phase to \emph{restrict the similarity condition} to recover more off-tree edges within one pass,
and \emph{parallelize disjoint subproblems} that avoid redundant work due to data dependencies as described in~\secref{prelim}.

\subsection{Density-Aware Approximation Condition}

The density-aware approximation restricts the similarity condition
in~\defref{similar-edge-loose} to require that \emph{both} endpoints of potentially similar edges be contained in the
separate $\beta$-hop neighborhoods of the
original edge endpoints rather than \emph{either} endpoint.
We further introduce a new definition for the parameter $\beta$ as follows.
\begin{equation}
    \beta_{u, v}^* = \min\{\text{dist}(u, \text{LCA}(u, v)),
                    \text{dist}(v, \text{LCA}(u, v)), c \} \label{eq:beta-new}
\end{equation}

More formally, the
density-aware condition replaces the \textit{OR} condition
with \textit{AND} condition in the endpoint membership.

\begin{definition}[Similiar edge (Strict)]\label{def:similar-edge-strict}
  Suppose we have a spanning tree $T$, an off-tree edge $e = (u, v)$, and
  BFS step size $\beta_{u, v}^*$ and the $\beta^*$-hop vertex sets
  $S_{u, \beta^*}, S_{v, \beta^*}$.
  An off-tree edge $e' = (u', v')$ is \defn{strictly similar} to off-tree edge
  $e$ if
  \begin{equation}
      (u' \in S_{u, \beta^*} \pmb{\wedge} v' \in S_{v, \beta^*})
      \vee (u' \in S_{v,\beta^*} \pmb{\wedge} v' \in S_{u, \beta^*}).
  \end{equation}
\end{definition}

\begin{figure}[t]
  \centering
  \includegraphics[width=\linewidth]{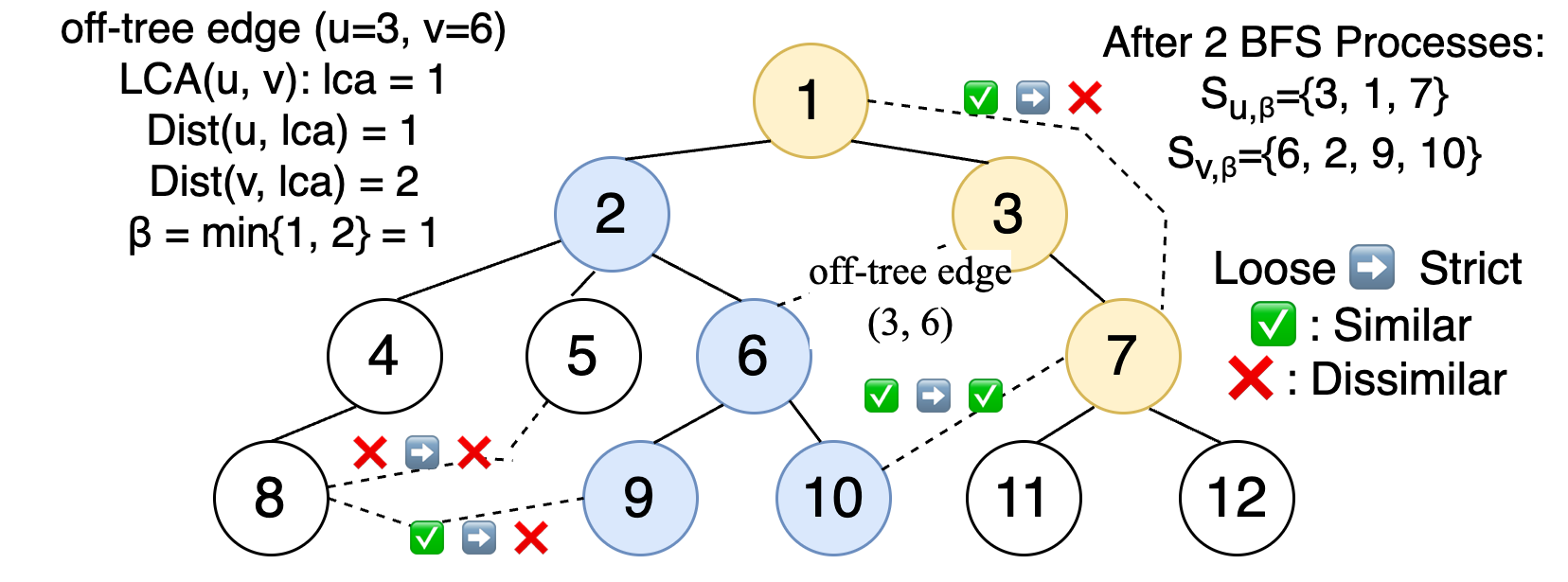}
  \caption{An illustration of the similarity check under the loose and strict similarity conditions. In
    this example, the edge being checked against is $e = (3, 6)$, $\operatorname{LCA}_T(3, 6) = 1$, and
    $\beta = 1$. The yellow (resp., blue) vertices denote $S_{u=3, \beta=1}$
    (resp., $S_{v=6, \beta=1}$). The arrow on the off-tree edges denotes whether the edge is loosely similar to $e$ (left of the arrow) and then whether the edge is strictly similar to $e$ (right of the arrow).
    }
    \label{fig:strict-condition}
\end{figure}

To clearly delineate between the two conditions going forward, we will use the
term \defn{loosely similar} to refer to the similarity definition in \fegrass and \pgrass (\defref{similar-edge-loose}).
~\figref{strict-condition} presents a worked example of edge similarity under the previous loose condition and the proposed strict condition.

Under strict similarity conditions, only two edges with highly similar positions
are considered similar, which enables \algnameshort to recover a sufficient number of
edges in a single recovery pass.
In contrast, \fegrass, which is based on vertex cover, tends to meet the
similarity condition more easily, and thus can only recover a small number
of edges in each pass.
In extreme cases, \fegrass may recover only one edge per round.
More specifically, if the constant $\beta$ is greater than or equal
to the diameter of the spanning tree, then recovering any single
edge can cover the entire graph.

\para{Properties of strictly-similar edges}
Next, we show how the strict edge-similarity condition in~\defref{similar-edge-strict}
give rise to disjoint subproblems without redundant work.
The full version of the paper will contain the proofs.

\begin{lemma}[Strictly similar edges share their LCA]\label{lem:similar-lca}
  If off-tree edge $e' = (u', v')$ is strictly similar to off-tree edge $e = (u,
  v)$, on a spanning tree $T$, then $\lca(u, v) = \lca(u', v')$.
\end{lemma}

\begin{lemma}[Contraposition]
  Given two off-tree edges $e = (u, v)$, $e' = (u', v')$ and a spanning tree $T$,
  if $\ \lca(u, v) \neq \lca(u', v')$, then $e'$ cannot be strictly similar to
  $e$. 
  \label{lem:shared-lca-contra}
\end{lemma}

\begin{lemma}[Strict similarity is non-commutative]
  \label{lem:similar-noncommutative}
  Given a spanning tree $T$, let $e = (u, v)$ and $e' = (u', v')$ be two
  strictly similar off-tree edges determined by first recovering $e$, performing
  BFS from $u$ and $v$ to get $\beta$-hop neighborhoods, and applying the strict
  similarity condition. If edge $e'$ was recovered first, edge $e$ may not have
  been marked strictly similar to $e'$.
\end{lemma}

~\lemref{similar-lca} proves that two strictly similar off-tree edges must share a LCA on spanning tree, which forms the basis for dividing all off-tree edges into groups based on their LCA.
Then ~\lemref{shared-lca-contra} (contraposition of ~\lemref{similar-lca}) forms the basis for dividing the off-tree edges into
\emph{disjoint subtasks} in \algnameshort based on the LCA of their endpoints. If the LCAs of the endpoints of two edges do not match, they cannot
possibly be strictly similar, so we can skip their similarity check.
Therefore, \algnameshort groups off-tree edges based on the LCA of their endpoints.
Only edges within a group need to be checked,  because edges cannot be strictly similar across groups.

~\lemref{similar-noncommutative} shows that the edges in the subtasks must be processed in sorted
order because strict similarity is non-commutative.
In other words, any parallelization within a subtask must process edges in order.
According to its definition in Equation~\ref{eq:beta-new}, $\beta$ is further constrained
by the minimum between a shorter distance and a predefined constant $c$, so the final value of $\beta$ is no greater than the
one used in the theoretical proof.
Hence, the introduction of the constant $c$ does not affect
the validity of the previously established properties.

Due to space limitations, we sketch some of the proofs in this section, but will include all proofs in the full version.

\section{Parallelizing \algnameshort}

This section will 1) show how the properties in~\secref{algorithm} give rise to
disjoint subtasks and  2) discuss several parallelization strategies
of the subtasks depending on their size.
Then we analyze the work and span of the proposed \algnameshort algorithm using the analysis model from~\secref{prelim}.

\subsection{Parallelization Strategies}

First, we propose several subtask parallelization strategies based on
the data dependencies shown in~\lemreftwo{shared-lca-contra}{similar-noncommutative}
and distribution of the size of the subtasks.

\para{Outer parallelization}
The most straightforward parallelization method is \defn{\outerpar parallelization}, or parallelization across the subtasks.
By~\lemref{shared-lca-contra}, all of these subtasks can be processed independently because there are no strictly similar edges across subtasks.
The similarity checks are embarassingly parallel across subtasks, so
there is no need for additional complexity.

\para{Inner parallelization}
If some subtasks are much larger than others, which is likely due to the skewed
nature of graphs, inter-task parallelization alone may not be sufficient to
achieve good parallel speedup.
Therefore, we can also perform \defn{\innerpar parallelization} within tasks
using the same blocked method as \pgrass described in~\secref{prelim}.
In the initial \pgrass parallelization method, the list of off-tree edges
is divided into blocks upfront, and edges in each block are processed in parallel,
but the blocks themselves are serialized.
Some of the edges in a block may have been marked during a previous block,
so the amount of edges that need to be processed in subsequent blocks
depends on previous blocks.

During each edge recovery, the algorithm first checks whether
the current edge has already been marked. If so,
it enters the continue branch and skips further processing.
When a set of edges is processed in parallel, threads that
enter the continue branch may lead to execution bubbles.
We propose an optimization method, \defn{Judge-before-Parallel}, that extracts the if-condition checks outside the parallel
region, thereby eliminating conditional branching and ensuring
full thread utilization without idle threads.

\para{Heuristic-based mixed parallel strategy}
In \algnameshort, we always apply outer-task parallelism across the independent subtasks,
but use heuristics to determine whether to apply inner-task parallelism in a given subtask based on whether the subtasks is sufficiently large (i.e., it has many edges or covers over 10\% of total edges).
We found that in practice, applying inner-task parallelism in subtasks with at least
\cutoff edges resulted in good parallel speedup.
Outer parallelism may encounter issues with uneven thread load.  When a very
large subtask is assigned to a particular thread, that thread may still be
executing even after all other subtasks have finished, which results in a
decline in overall performance.  Therefore, we need to apply an inner parallel
strategy for such large tasks.

\subsection{Algorithm Description and Analysis}

Finally, we describe \algnameshort 
and analyze its work and span in the binary-forking model described
in~\secref{prelim}. The full version will contain the proofs.

At a high level, \algnameshort applies the strict edge-similarity condition and
divides the edges into disjoint \defn{subtasks} according to the LCA of their
endpoints. From~\lemref{similar-lca}, if two edges are not strictly similar,
they have different LCAs. Edges must be processed sequentially within a subtask,
according to~\lemref{similar-noncommutative}, but \algnameshort can apply
inner parallel strategy of dividing the edges into blocks, as described
in~\secref{prelim}.

There are four main steps in \algnameshort: 1) computing the resistance distance
for each off-tree edge, 2) sorting the off-tree edges by their resistance
distance, 3) creating subtasks based on shared LCAs and sorting them based on
their size, and 4) recovering edges via the strict edge-similarity condition
within mixed parallel strategy on subtasks.
~\tabref{bounds-summary}
summarizes the work and span bounds for each of these steps.

\begin{table}[t]
\centering
 \begin{tabular}{l ccc}
 \hline
 Step & Work & Span\\ [0.5ex]
 \hline
 1 & $O(|E|\lg|V|))$ & $O(\lg^2|V|)$ \\
 2 & $O(|E|\lg|E|)$ & $O(\lg^2|E|)$  \\
3 &  $O(|E|\lg|E|)$ & $O(\lg^2|E|)$  \\
 4 & $O(\sum\limits^{k}_{i=0} |S_i|^2)$ & $O(|S_o|^2/p + |S_{\text{serial}}|^2)$
   \\
   Total & $O(|E|\lg|E| + \sum\limits^{k}_{i=0} |S_i|^2)$ & $O(\lg^2|E| +
                                                            |S_o|^2/p +
                                                            |S_{\text{serial}}|^2)$
   \\
 \hline
\multicolumn{3}{l}{
    $p$: number of parallel threads.
    $k$: number of subtasks.
}
\\
\multicolumn{3}{l}{
   $S_i$: set of edges in $i$-th subtask.
   $S_{\text{serial}}$: the largest sequential subtask.
}
\\
 \end{tabular}
 \caption{Work-Span analysis of each step in \algnameshort. }
 \label{tab:bounds-summary}
\end{table}


\begin{table*}
\setlength{\tabcolsep}{2.5pt}
\centering
  \resizebox{\textwidth}{!}{
    \begin{tabular}{
        l
        r r
        r r r r r r
        r r r r r r
        r r r r r r
    }

    \toprule
        & & 
        & \multicolumn{6}{c}{$\alpha = 0.02$} 
        & \multicolumn{6}{c}{$\alpha = 0.05$} 
        & \multicolumn{6}{c}{$\alpha = 0.10$} 
    \\
    \cmidrule(lr){4-9} \cmidrule(lr){10-15} \cmidrule(lr){16-21}
            Graph   & $|V|$     &  $|E|$
        & $T_{fe}$  & Pass  & $\iterfe$
            & $T_{pd-32}$   & $\iterpd$
            & $\frac{\iterfe}{\iterpd}$  
        & $T_{fe}$  & Pass  & \iterfe    
            & $T_{pd-32}$   & \iterpd   
            & $\frac{\iterfe}{\iterpd}$ 
        & $T_{fe}(ms)$  & pass  & \iterfe    
            & $T_{pd-32}$   & \iterpd  
            & $\frac{\iterfe}{\iterpd}$ 

    \\ 

    \midrule
    01-\textit{mi2010}
        & 3.30E5     & 7.89E5 
        & 82        & 1     &  \textbf{83}
            & \textbf{3}     &   93
            & 0.9
        & 116       & 3     &   66
            & \textbf{6}     &  \textbf{35}
            & 1.9
        & 180       & 6     &   57
            & \textbf{12}    &   \textbf{24}
            & 2.4
    \\
    02-\textit{mo2010}
        & 3.44E5     & 8.28E5 
        & 88        & 1     &  \textbf{84}
            & \textbf{4}     &   93
            & 0.9
        & 130       & 3     &   65
            & \textbf{6}     & \textbf{34}
            & 1.9
        & 202       & 6     &  57
            & \textbf{13}    &  \textbf{25}
            & 2.3
    \\
    03-\textit{oh2010}
        & 3.65E5     & 8.84E5 
        & 92        & 1     & \textbf{81} 
            & \textbf{3}      & 92
            & 0.9
        & 134       & 3     & 65
            & \textbf{8}      & \textbf{34}
            & 1.9
        & 210       & 6     & 53
            & \textbf{16}    &  \textbf{25}
            & 2.1
    \\
    04-\textit{pa2010}
        & 4.22E5     & 1.03E6 
        & 81        & 1     &  \textbf{83} 
            & \textbf{7}     & 93
            & 0.9
        & 117       & 3     & 66
            & \textbf{11}    & \textbf{34}
            & 1.9
        & 182       & 6     & 58 
            & \textbf{20}    &  \textbf{25}
            & 2.3
    \\    
    05-\textit{il2010}
        & 4.52E5     & 1.08E6 
        & 115        & 1     & \textbf{81}  
            & \textbf{11}      & 99
            & 0.8
        & 170       & 3     &  68
            & \textbf{13}      & \textbf{56}
            & 1.2
        & 270       & 5     & 55
            & \textbf{24}    & \textbf{26}
            & 2.1
    \\
    06-\textit{tx2010}
        & 9.14E5     & 2.23E6 
        & 276        & 1     & \textbf{87} 
            & \textbf{24}     & 97
            & 0.9
        & 471       & 3     &  72
            & \textbf{47}     & \textbf{34}
            & 2.1
        & 836       & 6     & 39
            & \textbf{96}    &  \textbf{27}
            & 1.4
    \\
    \hline
        07-\textit{com-DBLP}
        & 3.17E5     & 1.05E6 
        & 252        & 2     & \textbf{134}   
            & \textbf{118}     & 135
            & 1.0
        & 475       & 6     & 131 
            & \textbf{385}      & \textbf{121}
            & 1.1
        & \textbf{897}       & 14    & 124
            & 1139     & \textbf{105}
            & 1.2
    \\
        08-\textit{com-Amazon}
        & 3.35E5     & 9.26E5 
        & 130       & 2     & \textbf{72} 
            & 128     & 82
            & 0.9
        & \textbf{208}       & 4     & 61 
            & 347     & \textbf{60}
            & 1.0
        & \textbf{334}       & 7     & 54
            & 635    & \textbf{49}
            & 1.1
    \\
        09-\textit{com-Youtube}$^*$
        & 1.13E6     & 2.99E6 
        & -       & 6931     & 199   
            & \textbf{1353}      & \textbf{190}
            & 1.0
        & -       & 38308     & 224 
            & \textbf{8544}      & \textbf{127}
            & 1.8
        & -       & 103081     & 161
            & \textbf{23062}     & \textbf{93}
            & 1.7
    \\
    \hline
        10-\textit{coAuthorsCiteseer}
        & 2.27E5     & 8.14E5 
        & 192       & 2     &   175
            & \textbf{53}     &  \textbf{111}
            & 1.6
        & 341      & 5     &  160
            & \textbf{160}      & \textbf{101}
            & 1.6
        & 582       & 10      & 128
            & \textbf{387}     & \textbf{93}
            & 1.4
    \\
        11-\textit{citationsCiteseer}
        & 2.68E5     & 1.16E6 
        & 514       & 4     &  131  
            & \textbf{115}     & \textbf{126}
            & 1.0
        & 1245      & 15     & \textbf{94} 
            & \textbf{338}    & 113
            & 0.9
        & 3021       & 52     & 104
            & \textbf{791}    & \textbf{70} 
            & 1.5
    \\    
        12-\textit{coAuthorsDBLP}
        & 2.99E5     & 9.78E5 
        & 248       & 3     & \textbf{86}   
            & \textbf{105}       & 89
            & 1.0
        & 477      & 6     & 83 
            & \textbf{342}       & \textbf{74}
            & 1.1
        & 883       & 14     & 80
             & 889      & \textbf{62}
             & 1.3
    \\    
        13-\textit{coPapersDBLP}
        & 5.40E5     & 1.52E7 
        & 10770       & 2     & \textbf{200}  
            & \textbf{105}       & 234
            & 0.9
        & 22813      & 6     & \textbf{206} 
            & \textbf{420}       & 215
            & 1.0
        & 42865       & 13     & 202
            & \textbf{3703}      &  \textbf{192}
            & 1.1
    \\
    \hline
        14-\textit{NACA0015}
        & 1.04E6     & 3.11E6 
        & 352      & 1     & \textbf{98}  
            & \textbf{26}       & 234
            & 0.4
        & 519      & 2     & \textbf{83} 
            & \textbf{49}       & 85
            & 1.0
        & 834       & 4     & 66
            & \textbf{92}      & \textbf{30}
            & 2.2
    \\
        15-\textit{M6}
        & 3.50E6     & 1.05E7 
        & 1500      & 1     &  \textbf{105} 
            & \textbf{164}      & 106
            & 1.0
        & 2192      & 2     & \textbf{89} 
            & \textbf{372}      & 91
            & 1.0
        & 3520       & 4     & 66
            & \textbf{786}     & \textbf{31}
            & 2.1
    \\    
        16-\textit{333SP}
        & 3.71E6     & 1.11E7 
        & 1550      & 1     & \textbf{107}  
            & \textbf{222}      & 187
            & 0.6
        & 2218      & 2     & 88 
            & \textbf{419}      & \textbf{85}
            & 1.0
        & 3489       & 4     & 69
            & \textbf{865}     & \textbf{30}
            & 2.3
    \\    
        17-\textit{AS365}
        & 3.80E6     & 1.14E7 
        & 1629      & 1     & \textbf{105} 
            & \textbf{242}      & 182
            & 0.6
        & 2390      & 2     & \textbf{87} 
            & \textbf{437}      & 91
            &  1.0
        & 3863       & 4     & 70
            & \textbf{906}     & \textbf{30}
            & 2.3
    \\    
        18-\textit{NLR}
        & 4.16E6     & 1.25E7 
        & 1818      & 1     & \textbf{108}  
            & \textbf{221}      & 193
            & 0.6
        & 2674      & 2     & \textbf{89}
            & \textbf{502}      & 90
            & 1.0
        & 4256       & 4     & 71
            & \textbf{1052}     & \textbf{30}
            & 2.4
    \\
      \bottomrule
      
    \multicolumn{21}{l}{ 
        $*$Time out for \fegrass on graph \textit{com-Youtube}, for $\alpha = 0.02, 0.05$, \fegrass runs over 10 minutes; 
        for $\alpha = 0.10$, \fegrass runs over 1 hour. 
    } 
    \end{tabular}
}
    \caption{
        Evaluation of runtime and subgraph quality. 
        $T_{fe}$ (ms) and $T_{pd-32}$ (ms) represent the execution times of \fegrass and \pdgrass (with 32 threads), respectively.
        $Pass$ represents the number of passes required by \fegrass to restore a sufficient number of edges. 
        The $Pass$ for \algnameshort is omitted because it always completes in one single pass.
        \iterfe and \iterpd represents the number of iterations
        for the PCG solver to converge when using the generated sparsifier as the preconditioner. 
        }
  \label{tab:parallel_scaling}
 \vspace{-.5cm}
\end{table*}

\section{Evaluation}\label{sec:eval}

This section empirically evaluates \algnameshort compared to \fegrass
in terms of their 1) recovery runtime
and 2) sparsifier quality, in terms of the iteration count required for convergence
when using the sparsifier in a downstream application of
preconditioned conjugate gradient (PCG) ~\cite{barrett1994templates}.
Then we evaluate the strong scaling of \pdgrass on representative cases of skewed and uniform subtask distributions to evaluate the impact of different parallelization strategies.
Experiments show that the mixed parallel strategy consistently achieves effective parallel scaling across diverse data distributions.

\para{Setup} We performed experiments on an Rocky Linux (8.9, Green Obsidian)
server with AMD EPYC 7T83 64-Core Processor.  Each core has 32 KB L1d/L1i and
512 KB L2 caches, with a 32 MB L3 cache shared per socket.  We implemented
\algnameshort in \texttt{C++17} with \texttt{OpenMP 4.5}~\cite{openmp15} and
compiled by \texttt{GCC 8.5.0}.  We compare with the
open-source\footnote{https://github.com/5Mrzhao/CSP/blob/main/Sfegrass.cpp}
implementation of \fegrass.  To do an apples-to-apples comparison of the
recovery step, \algnameshort uses the same spanning tree as \fegrass.  We
report the minimum \emph{runtime} over 5 trials of the edge recovery step.

The parameter $\alpha$ defines the ratio of edges to be recovered.
We evaluate \fegrass and \algnameshort with $\alpha$ set to 0.02, 0.05, and 0.10 to assess performance across different recovery ratios.

We measure sparsifier \emph{quality} by the iteration count
required for convergence in a PCG solver from MATLAB R2023b to evaluate GSS performance.
Specifically, given a subgraph $P$ of the original graph $G$,
the PCG solver uses $L_P$ as the preconditioner
to solve $||L_Gx - b|| \le 10^{-3}||b||$ iteratively.
A lower iteration count indicates a higher-quality sparsifier.

\para{Datasets} We evaluate the algorithms on a suite of \numgraphs graphs from
the SuiteSparse Matrix Collection~\cite{Matrixset}, including datasets from
SNAP~\cite{snapnets} and DIMACS10~\cite{DIMACS10}.  We select symmetric matrices
representing undirected graphs with a single connected component.  For graphs
without edge weights, we assign random positive weights uniformly sampled
between 1 and 10.  ~\tabref{parallel_scaling} summarizes the graph
sizes.

\para{\algnameshort parameters}
Given $p$ threads, we set the block size (as defined in~\secref{prelim}) to $p$,
enabling $p$ edges to be processed in parallel.
This design follows the \textit{Judge-Before-Parallel} optimization
to maximize thread occupancy.
\algnameshort first handles large tasks using \innerpar parallelism in a one-by-one manner and then applies \outerpar parallelism to the remaining tasks.
The cutoff between \innerpar and \outerpar is defined as \cutoff, or 10\% of the total off-tree edges, effectively isolating larger subtasks for improved performance.
As for $\alpha$, we choose $0.02, 0.05, 0.10$ and report the results in~\tabref{parallel_scaling}. The default in previous work~\cite{liu2021fegrass} is $\alpha = 0.02$, but we show that \algnameshort with increasing $\alpha$ can efficiently compute higher-quality subgraphs than with smaller $\alpha$.

\para{Runtime}
On 32 threads, \algnameshort shows a significant runtime advantage over \fegrass.
Excluding the \textit{com-youtube} dataset due to a timeout,
\algnameshort achieves average speedups of $8.76\times$, $6.32\times$, and $3.94\times$
over \fegrass under 32-thread execution,
for $\alpha = 0.02$, $0.05$, and $0.10$, respectively.
These results align with our algorithmic analysis:
as $\alpha$ increases and more edges are recovered, the problem size grows,
causing the runtime of \algnameshort with single thread to increase faster
than that of \fegrass due to its inherently quadratic complexity,
whereas \fegrass operates in linear time.

As discussed in~\secref{intro}, \algnameshort addresses the worst-case
runtimes encountered by \fegrass on challenging inputs.
Although \fegrass has lower theoretical complexity, its vertex-cover-based loose
similarity condition leads to many edges being marked as similar,
often requiring multiple passes to meet recovery targets.
\textit{com-Youtube}, a highly skewed graph where a few high-degree vertices connect
to many others, exemplifies this issue. Once a high-degree vertex is covered,
nearly all incident edges are marked as similar, severely limiting the number
of edges recoverable per pass. Even with $\alpha = 0.02$, \fegrass requires
over 6000 passes to recover enough edges. While each pass runs in linear time,
the cumulative overhead results in significant performance degradation.
Excluding \textit{com-youtube}, \fegrass still requires an average
of 1.6, 4.1, and 9.7 passes to recover 2\%, 5\%, and 10\% of edges, respectively.
In contrast, \algnameshort’s stricter similarity condition allows substantially
more edges to be recovered per pass.

\para{Quality}
As shown in~\tabref{parallel_scaling}, \algnameshort yields
lower PCG iteration counts than \fegrass on most datasets as $\alpha$ increases,
indicating improved subgraph quality.
The iteration ratio between \fegrass and \pdgrass rises from $0.9\times$
at $\alpha = 0.02$ to $1.3\times$ at $\alpha = 0.05$, and $1.8\times$
at $\alpha = 0.10$, suggesting that \pdgrass benefits more from increased edge recovery.
At $\alpha = 0.10$, \algnameshort consistently achieves convergence
with roughly half the PCG iterations required by \fegrass.
Although $\alpha = 0.02$ is commonly used in \fegrass as a heuristic
to balance quality and runtime, \algnameshort’s higher efficiency
enables the use of larger $\alpha$ values to produce higher-quality subgraphs
without significant overhead.

The strict similarity condition in \pdgrass
allows more spectrally critical edges to be recovered,
leading to higher-quality subgraphs.
In contrast, \fegrass employs a looser condition,
which often causes some spectrally important edges to be marked
as similar and excluded from recovery once their endpoints are covered.


\para{Strong Scalability}
We conducted a comparison between \fegrass and \algnameshort by evaluating their
performance ratios on 1, 8, and 32 threads.  As the number of threads increases
to 32, \algnameshort consistently surpasses \fegrass in performance across all
datasets, with an average speedup of $8.8 \times$.  Notably,
\algnameshort delivers over $20 \times$ speedup on 5 out of 18 tested datasets.  
Furthermore, \algnameshort scales efficiently, achieving an
average speedup of $5.8 \times$ on 8 threads and $11.3 \times$ on 32 threads.
On skewed inputs, the inner parallel region dominates execution time due to
extreme load imbalance, while the outer region contributes minimally and quickly
saturates in speedup.  A mixed parallel strategy is essential in this case to
balance the workload and improve scalability.  On more uniform inputs, even task
distribution enables efficient parallelism, resulting in near-ideal strong
scaling across threads.
More detailed tables will be included in the full version.


\section{Conclusion}

We present \algnameshort, an efficient parallel Graph Spectral Sparsification (GSS)
algorithm generating density-aware subgraphs.
\algnameshort improves the state-of-the-art \fegrass
by 1) restricting edge-similarity conditions to recover more edges per pass,
and 2) parallelizing over the resulting independent subtasks.
Experiments demonstrate significant enhancements in graph quality and runtime.
A mixed parallel strategy is proposed to handle graph structures
with diverse subtask distributions, ensuring balanced workload and full thread utilization.

\clearpage
\balance
\bibliographystyle{IEEEtran}
\bibliography{references}

\newpage
\appendices

\section{Proofs}

\begin{figure}[t]
  \centering
  \includegraphics[width=0.65\linewidth]{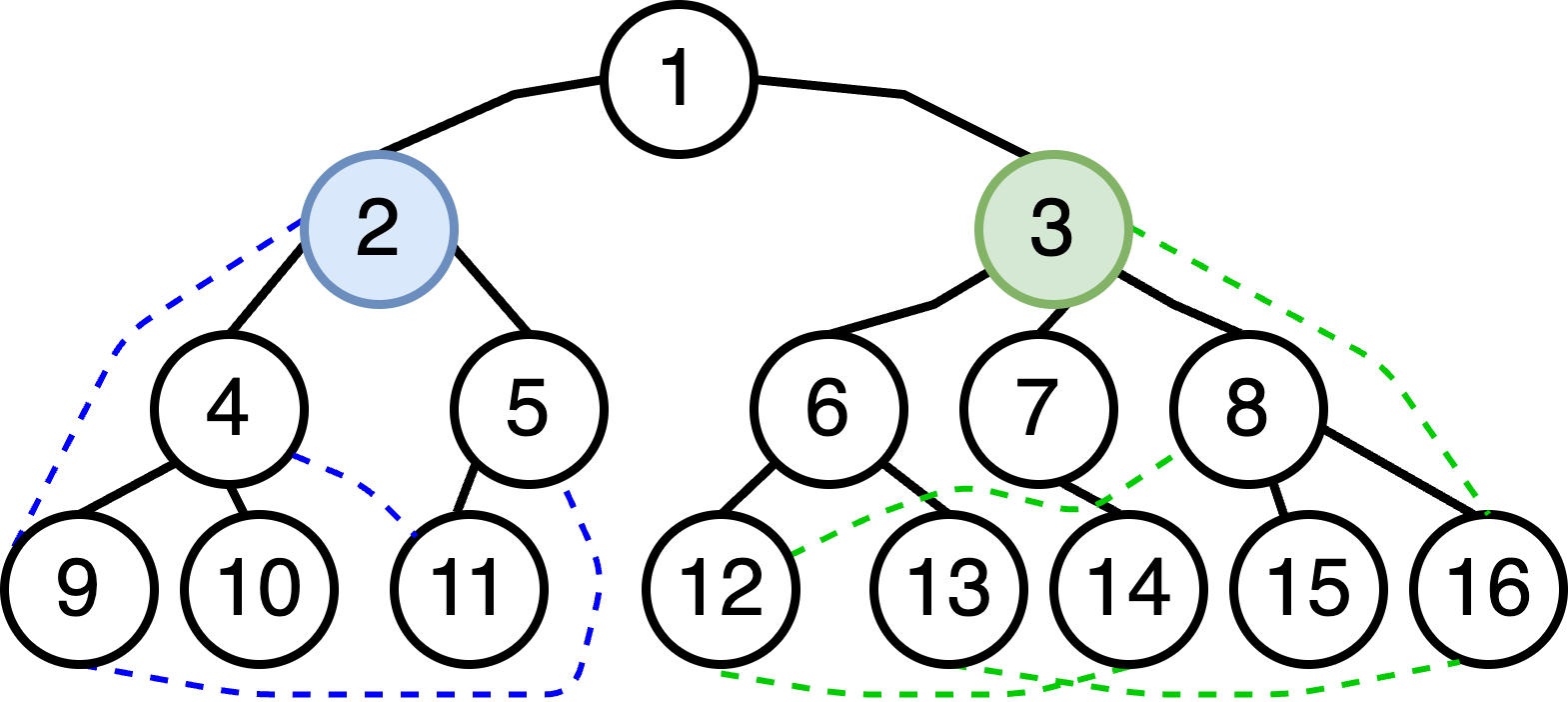}
  \caption{An example of subtask division based on shared LCAs. The endpoints of
    edges $(2, 9), (4, 11), (5, 9)$ shares the same LCA node (2), so these edges
    fall into same (blue) subtask.
    Similarly, edges $(3, 16)$, $(8,12)$, $(12, 14)$, $(13, 16)$
    share the same LCA node (3) and fall into green subtask.
    Blue and green edges cannot be strictly
    similar to each other, so the subtasks are independent.}
    \label{fig:lca-subtask-division}
\end{figure}

\begin{figure}[t]
  \centering
  \includegraphics[width=1.0\linewidth]{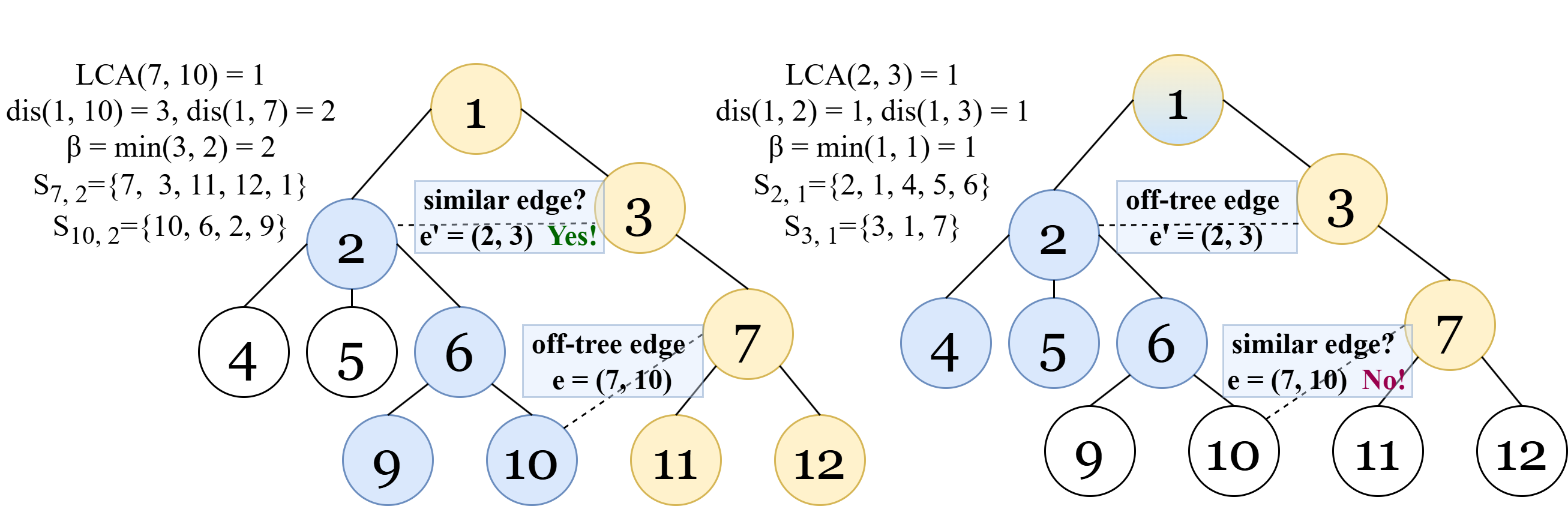}
  \caption{Non-commutative property of strict similarity.
    On the left side, edge $e = (7, 10)$ is recovered first,
    and edges $e' = (2, 3)$ is marked strictly similar to $e$.
    On the right side, if $e'$ is recovered first,
    $e$ is \emph{not} marked strictly similar to $e'$.  }
\label{fig:noncommutative}
\end{figure}

\quad \\
\textbf{Lemma} (Strictly similar edges share their LCA)
  If off-tree edge $e' = (u', v')$ is strictly similar to off-tree edge $e = (u,
  v)$,
  on a spanning tree $T$, then $\lca(u, v) = \lca(u', v')$. 
\\
\textbf{Lemma} (Contraposition)
  Given two off-tree edges $e = (u, v)$, $e' = (u', v')$ and a spanning tree $T$,
  if $\ \lca(u, v) \neq \lca(u', v')$, then $e'$ cannot be strictly similar to
  $e$.

\begin{proof}
  Without loss of generality, let $u' \in \su$ and $v' \in \sv$. By
  construction, $\lca(u, v)$ is also an ancestor of both $u'$ and $v'$.  For
  contradiction, suppose there is another vertex
  $w = \lca(u', v') \neq \lca(u, v)$, such that $w$ has greater depth than
  $\lca(u, v)$. Therefore,
  \begin{align*}
    \dist(u', v') & = \dist(u', w) + \dist(w, v') \\
                    & < \dist(u', \lca(u, v)) +
                      \dist(\lca(u, v), v').
  \end{align*}
  However, there is only one path between any two vertices on a tree. If there
  were another such vertex $w$ with greater depth, it would have also been the
  LCA of $u$ and $v$. Thus, the LCA for both pairs of vertices must be the
  same: $\lca(u', v') = \lca(u, v)$.
\end{proof}

~\\
\textbf{Lemma} (Strict similarity is non-commutative)
  Given a spanning tree $T$, let $e = (u, v)$ and $e' = (u', v')$ be two
  strictly similar off-tree edges determined by first recovering $e$, performing
  BFS from $u$ and $v$ to get $\beta$-hop neighborhoods, and applying the strict
  similarity condition. If edge $e'$ was recovered first, edge $e$ may not have
  been marked strictly similar to $e'$.

\begin{proof}
  Without loss of generality, assume $\dist(\lca, u) \leq \ \ $
  $\dist(\lca,v)$, so $\beta = \dist(\lca, u)$.
  Also without loss of generality, suppose that
  $u' \in \su$ and $v' \in \sv$, and that
  $\beta' = \dist(\lca, v') < \dist(\lca, u')$.
  Suppose that $\beta' < \beta$. We will proceed with proof by construction. Let
  $\beta' < \dist(\lca, v)/2$. Therefore, $v'$ is $v$'s ancestor and \lca's
  descendant. By definition, $\dist(\lca, v) = \dist(\lca, v') + \dist(v' v)$
  and $\dist(lca, v') < dist(v', v)$.  Since
  $\beta' = \dist(\lca, v') < dist(v', v)$, when performing $\beta'$-hop BFS
  from vertex $v'$, vertex $v$ will not be visited.  Thus, off-tree edge $e$
  would not be marked similar to $e'$ if $e'$ was recovered first.
  ~\figref{noncommutative} illustrates an example of non-commutative property for strict similarity.
\end{proof}

~\\
\textbf{Lemma} (Work analysis)
  Let $S_i$ denote the set of edges in subtask $i$, and let $k$ be the number of subtasks.
  The \algnameshort algorithm has $O(|E|\lg|E| + \sum\limits^{k}_{i=0} |S_i|^2)$ work.

\begin{proof}
  The first step computes the resistance distance for each off-tree edge, which
  takes $O(|E|\lg|V|)$ work using binary lifting as described earlier.
  The second step sorts the off-tree edges based on their resistance
  distance, which takes $O(|E|\lg|E|)$ work with any standard sorting algorithm.
  Preparing the subtasks takes $O(|E|\lg|E|)$ work with one serial pass through
  the off-tree edges and subsequent sorting of subtasks based on their size in
  edges.
  The fourth step recovers off-tree edges according to the strict
  edge-similarity condition. In the worst case, every edge in each subtask must
  check against every edge after it in the subtask if no edges are marked
  similar. Therefore, the worst-case work of subtask $i$ is $|S_i|^2$, and the
  worst-case work over all the subtasks is
  $O(|E|\lg|E| + \sum\limits^{k}_{i=0} |S_i|^2)$.
  The work of the entire algorithm is $O(|E|\lg|E| + \sum\limits^{k}_{i=0}
  |S_i|^2)$ by adding up the work of the steps.
\end{proof}

~\\
\textbf{Lemma} (Span analysis)
  Let $p$ denote the number of parallel threads, $S_i$ denote the set of edges
  in subtask $i$, and let $S_{\text{serial}}$ denote the largest subtask that is
  performed sequentially. The \algnameshort algorithm has
  $O(\lg^2|E| + |S_0|^2/p + |S_{\text{serial}}|^2)$ span.

\begin{proof}
  The span of the first step is $O(\lg^2|V|)$ from the span of constructing the
  skip table. The $|E|$ LCA and distance queries have $O(\lg|E| + \lg|V|)$ span,
  which is asymptotically smaller than the span of skip-table construction.

  Steps 2 and 3 can be solved with any standard parallel sorting algorithm, such
  as parallel merge sort, in $O(\lg^2|E|)$ span~\cite{CLRS}.

  The span of the edge-recovery step is $O(|S_o|^2/p +
  |S_{\text{serial}}|^2)$. All of the subtasks can proceed in parallel, and some
  of them are further parallelized internally using the same blocking method as
  \pgrass. The largest task $S_o$ dominates the time for the tasks that use
  intra-task parallelism, and processing it takes $O(|S_0|^2/p)$ span. In the
  worst case, no edges are marked as similar, so the subtask takes quadratic
  work, but all of the comparisons can proceed in parallel. Furthermore, the
  largest serial task dominates the span for the sequential tasks, and
  contributes $O(|S_{\text{serial}}|^2)$ to the span in the worst case.
  The span of the entire algorithm is
  $O(\lg^2|E| + |S_o|^2/p + |S_{\text{serial}}|^2)$ by adding up the span of the
  steps.
\end{proof}

\section{Algorithm details}

\begin{algorithm}[t]
\caption{\algnameshort}\label{alg:pdgrass}
\KwData{$\texttt{L}$ = list of off-tree edges, $\texttt{T}$ = maximum spanning tree}
\KwResult{\texttt{L2} = list of recovered off-tree edges}
Let \texttt{dists} = list of the distances of each edge's endpoints to $\texttt{T}$'s
root \\
\tcc{Step 1 - Precompute skip table for LCA }
Generate skip table $\texttt{S}$ to answer LCA queries efficiently \\
\ParallelFor{$i \in [0, \texttt{num\_off\_tree\_edges})$}{
  \tcc{Compute resistance distance}
  \texttt{L[i].lca} = \texttt{LCA\_query(L[i], S)} \\
  \texttt{dists[i]} = \texttt{distance\_query(L[i], S)} \\
  \texttt{L[i].resistance} = \texttt{compute\_resistance(L[i], dists[i])}
}
\tcc{Step 2 - Sort \texttt{L}}
Parallel stable sort \texttt{L} based on resistances \\
\tcc{Step 3 - Create subtasks} 
Create index-based linked lists of subtasks in \texttt{L} 
\\
Sort subtasks on number of edges in subtask\\
\tcc{Step 4 - Process subtasks}
Split subtasks into $S_{large}, S_{small}$ with \texttt{CUTOFF} \\
\For {$s_i \in S_{large}$} {
    Apply \textit{Inner Parallel Strategy} on $s_i$ 
}
\ParallelFor{$s_i \in S_{small}$}{
    Parallel process edge recovery on $s_i, ..., s_{i+Thread\_cnt}$
}
\Return{\texttt{L2}}
\end{algorithm}

\subsection{Analysis}

\para{Work analysis}
First, we give an analysis of the work in the worst case, which is dominated by
the edge-recovery phase and therefore the subtasks. Theoretically, the
worst-case work is quadratic in the size of the subproblems because the
algorithm cannot guarantee that any edges marked as similar, so all edges must
be compared with all other edges. However, in practice, the runtime is much
better than the suggested worst-case bound, because many edges are marked
similar and can be skipped.

\begin{lemma}[Work analysis]
  Let $S_i$ denote the set of edges in subtask $i$, and let $k$ be the number of subtasks.
  The \algnameshort algorithm has $O(|E|\lg|E| + \sum\limits^{k}_{i=0} |S_i|^2)$ work.
\end{lemma}

\para{Span analysis}
Next, we analyze the span, or the longest path through the computation DAG (and
the time on infinitely many processors) in the binary forking model described
in~\secref{prelim}.

\begin{lemma}[Span analysis]
  Let $p$ denote the number of parallel threads, $S_i$ denote the set of edges
  in subtask $i$, and let $S_{\text{serial}}$ denote the largest subtask that is
  performed sequentially. The \algnameshort algorithm has
  $O(\lg^2|E| + |S_0|^2/p + |S_{\text{serial}}|^2)$ span.
\end{lemma}


\section{\textit{Judge-Before-Parallel} Optimization}

\begin{table}[t]
\centering
\begin{tabular}{l l l}
\toprule
   \textit{Graph: com-youtube}     & $Without$  & $With$  \\
\midrule
\# off-tree edges in biggest task   & 1852735   & 1758087   \\
\# edges in parallel blocks               & 1852735                    & 719602 \\
\# edges skipped in parallel            &  1058401 (57\%)   & 0 \\
\# edges explored in parallel        &  794334  (43\%)   & 719602 (100\%)  \\
\# false positive edges           &  3048
& 60 \\ 
\bottomrule
\end{tabular}
\caption{Statistics of \algnameshort on the com-youtube graph with and without the \textit{Judge-Before-Parallel} optimization.}
\label{tab:judge_table}
\end{table}


~\tabref{judge_table} shows that on the skewed com-youtube graph,
where $95\%$ of the edges are in the largest subtask, 57\% of
thread iterations enter the ``continue'' branch during execution
and become idle without \textit{Judge-before-Parallel} optimization.
For the other $43\%$ of thread iterations, $0.4\%$ of the computation
is redundant which does not contribute to the final results.
In contrast, the inner parallel strategy in \algnameshort with
the  \textit{Judge-before-Parallel} optimization
reduces the percentage of redundant edge computations to $0.01\%$.
\section{Data}

\begin{table}[t]
\centering
\setlength{\tabcolsep}{3pt}
\small  
\begin{tabular}{
    l
    r 
    r r 
    r r 
    r r
    r
}
\toprule
 
   \textit{Graph} & $T_{fe}$   
    & $T_{1}$    & $ \frac{T_{fe}}{T_{1}}$    
    & $T_{8}$    & $ \frac{T_1}{T_{8}} $ 
    & $T_{32}$   & $\frac{T_1}{T_{32}}$
    & $\frac{T_{fe}}{T_{32}}$    
\\
\midrule
01   
    & 82
    & 58    & 1.4
    & 7     & 8.3
    & 3     & 19.3
    & 27.3
    \\
02      
    & 88
    & 54    & 1.6
    & 9     & 6.0
    & 4     & 13.5
    & 22.0
    \\
03      
    & 92
    & 58    & 1.6
    & 10     & 5.8
    & 3     & 19.3
    & 30.7
    \\
04   
    & 81
    & 91    & 0.9
    & 13     & 7.0
    & 7     & 13.0
    & 11.6
    \\
05        
    & 115
    & 100    & 1.2
    & 15     & 6.7
    & 11     & 9.1
    & 10.5
    \\
06       
    & 276
    & 350    & 0.8
    & 53     & 6.6
    & 24     & 14.6
    & 11.5
    \\
\hline
07     
    & 252
    & 762     & 0.3
    & 200     & 3.8
    & 118     & 6.5
    & 2.14
    \\
08   
    & 130
    & 540     & 0.2
    & 157     & 3.4
    & 128     & 4.2
    & 1.0
    \\
09$^*$
    & - 
    & 10631    & - 
    & 2409     & 4.4
    & 1353     & 7.9
    & -
    \\
\hline
10 
    & 192
    & 207    & 0.9
    & 68     & 3.0
    & 53     & 3.9
    & 3.6
    \\
11 
    & 514
    & 592    & 0.9
    & 185       & 3.2
    & 115       & 5.2
    & 4.5
    \\
12     
    & 248
    & 635    & 0.4
    & 170     & 3.7
    & 105     & 6.1
    & 2.4
    \\
13      
    & 10770
    & 748    & 14.4
    & 180     & 4.2
    & 105     & 7.1
    & 102.6
    \\
\hline
14     
    & 352
    & 469    & 0.8
    & 57     & 8.2
    & 26     & 18.0
    & 13.5
    \\
15           
    & 1500
    & 4528    & 0.3
    & 410     & 11.0
    & 164     & 27.6
    & 9.2
    \\
16        
    & 1550
    & 4157    & 0.4
    & 547     & 7.6
    & 222     & 18.7
    & 7.0
    \\
17       
    & 1629
    & 5326    & 0.3
    & 551     & 9.7
    & 242     & 22.0
    & 6.7
    \\
18          
    & 1818
    & 6080    & 0.3
    & 628     & 9.7
    & 221     & 27.5
    & 8.2
    \\
\bottomrule
\end{tabular}
\caption{Runtime(ms) of \fegrass (1 thread) and \algnameshort on $1/8/32$ threads ($T_1$, $T_8$, and $T_{32}$, respectively), when $\alpha=0.02$.}
\label{tab:scalability}
\vspace{-.1cm}
\end{table}

\subsection{Parallel Scalability of \algnameshort}

\begin{figure}[t]
        \centering
        \includegraphics[width=0.8\linewidth]{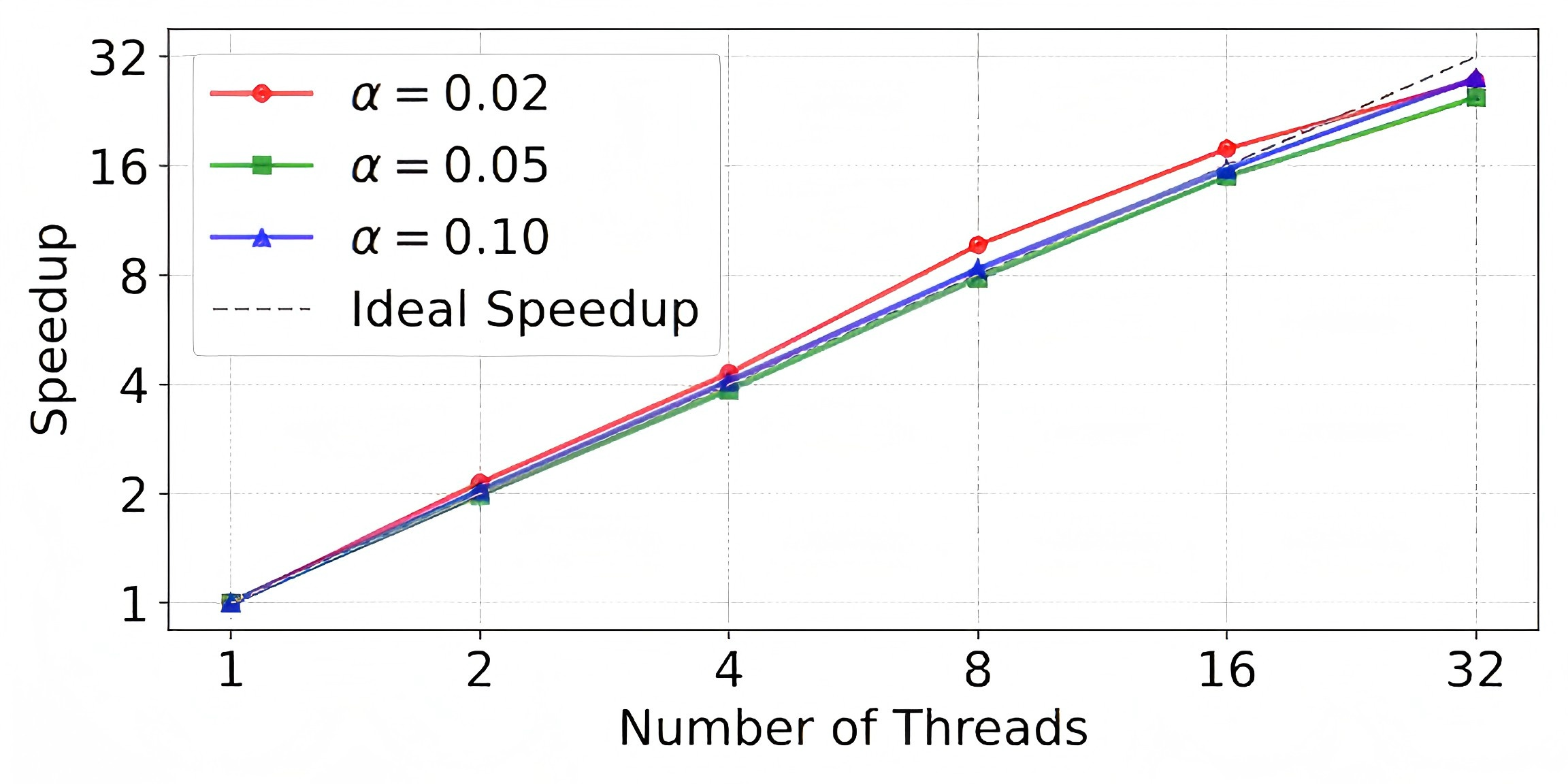}
        \captionsetup{skip=3pt}
        \caption{Strong scaling (Entire Outer Parallel) on \textit{M6}. }
        \label{fig:speedup-m6-outer}
\end{figure}

\begin{figure}[t]
        \centering
        \includegraphics[width=0.8\linewidth]{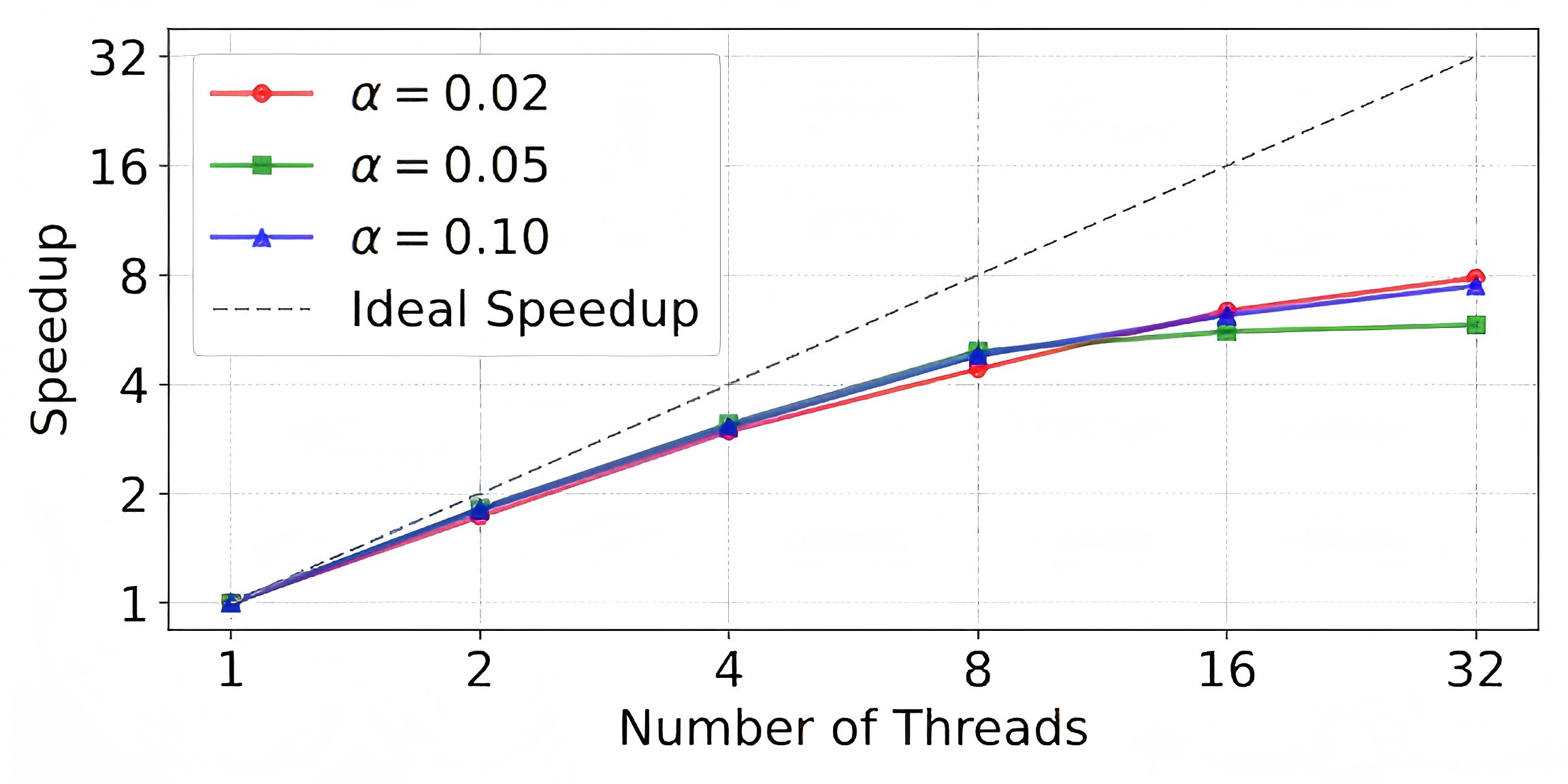}
        \captionsetup{skip=3pt}
        \caption{Strong scaling (Inner Parallel Part) on \textit{com-Youtube}. }
        \label{fig:speedup-yt-inner}
\end{figure}

\begin{figure}[t]
        \centering
        \includegraphics[width=0.8\linewidth]{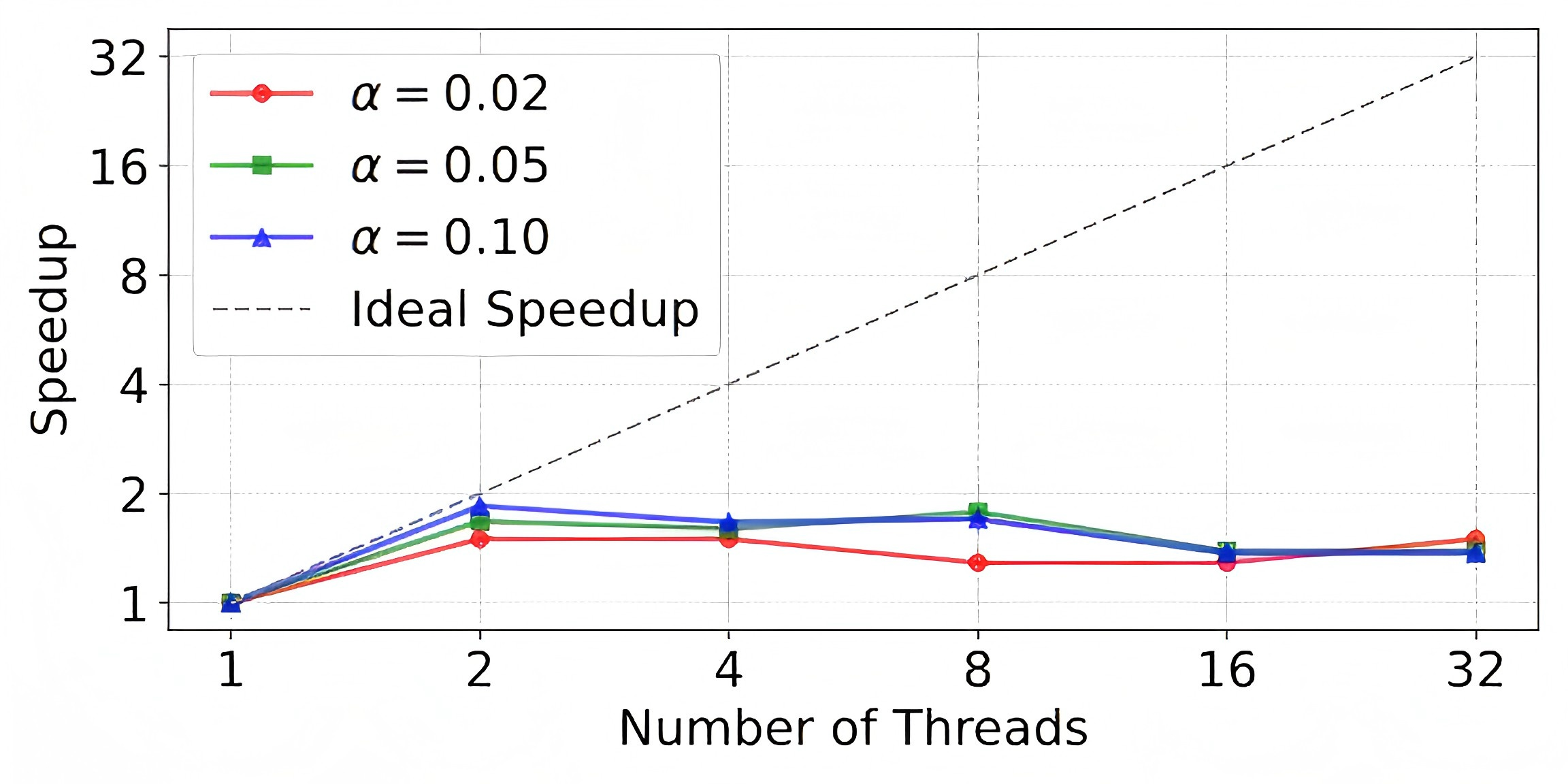}
        \captionsetup{skip=3pt}
        \caption{Strong scaling (Outer Parallel Part) on \textit{com-Youtube}. }
        \label{fig:speedup-yt-outer}
\end{figure}

Next, we measure 1) the scalability of \algnameshort on all datasets with 8 and 32 threads as well as the 2) strong scaling with increasing thread count on two representative inputs. Due to space limitations, we report results with
with $\alpha = 0.02$ in~\tabref{scalability}, but we find that the results are similar with different $\alpha$.


First, we conducted a comparison between \fegrass and \algnameshort
by evaluating their performance ratios on 1, 8, and 32 threads. In a single-threaded setting,
\fegrass demonstrates better performance than \algnameshort across
the majority of datasets, aligning with our prior theoretical
complexity analysis. As the number of threads increases to 32,
\algnameshort consistently surpasses \fegrass in performance across
all datasets, with an average speedup of $8.8 \times$. Notably,
\algnameshort delivers over $20 \times$ speedup on many of the tested datasets.
Furthermore, \algnameshort scales efficiently, achieving an average speedup of $5.8 \times$ on 8 threads
and $11.3 \times$ on 32 threads.

\para{Strong scaling on representative cases} Furthermore, we select two representative cases to further
investigate typical subtask distribution patterns.
The \textit{com-Youtube} graph represents a skewed distribution,
where the largest subtask occupies the majority of the workload.
In contrast, the \textit{M6} dataset illustrates a uniform
distribution, characterized by a large number of subtasks,
each with relatively small data sizes.
These two representative subtask distributions reflect two
extreme cases of parallel execution strategies within the algorithm.
In scenarios with a skewed distribution, the largest subtask
accounts for most of the workload, rendering the
outer parallelism ineffective.
In such cases, parallel efficiency relies heavily on
inner parallel execution.
In contrast, when subtasks are uniformly distributed,
the algorithm relies entirely on outer parallelism,
taking advantage of the balanced workload to deliver
strong scalability and improved performance.

\para{Scaling on skewed inputs}
First, we study the strong scaling on the skewed com-Youtube graph.  In this case, the inner parallel part accounts for nearly the entire
execution time because the largest subtask accounts for over 99\%
of the off-tree edges, leading to severe load imbalance across threads. Therefore, a mixed parallel strategy becomes
essential to effectively handle such workload imbalances.

~\figreftwo{speedup-yt-inner}{speedup-yt-outer}
show the thread speedup for inner and outer parallel parts, respectively.
In contrast, the outer parallel component processes a limited number
of subtasks with relatively small data volumes, contributing only
a negligible portion of the total runtime.
For outer parallel part, tasks with fewer subtasks and smaller scales
quickly reach an upper bound in speedup as the number of threads
increases. In~\figref{speedup-yt-outer}, a good speedup is
achieved at 2 threads, but as the thread count continues to rise,
the speedup plateaus at around $2\times$.
For inner parallel part, although the algorithm introduces additional
computations for serial checks to ensure correctness,~\figref{speedup-yt-inner} still shows that the speedup continues
to increase as the number of threads grows, achieving
around $8 \times$ speedup with 32 threads.

\para{Scaling on uniform inputs}
Next, we measure the strong scaling on the M6 graph, which exhibits a more uniform distribution on the subtask sizes. In this case, the largest subtask accounts for less than 0.3\%,
while the number of subtasks is large and their distribution
is relatively balanced.
In this scenario, all tasks are assigned to the outer parallel part,
as this ensures even thread load distribution and
keeps threads fully utilized.
As shown in~\figref{speedup-m6-outer}, the speedup approaches
the ideal value with thread scaling (about $28\times$ speedup on 32 threads).

\end{document}